\documentclass[]{emulateapj}%{aastex}
\usepackage{graphicx} 
\usepackage{ulem}
\usepackage{rotating}
\usepackage{lscape}

\newcommand{\kms}{km~s$^{-1}$}

\newcommand{\sci}[1]{{\rm \; \times \; 10^{#1}}}

\newcommand{\Ha}{H$\alpha$}

\begin{document}

% %for title footnote
%\makeatletter
%\def\@fnsymbol#1{\ensuremath{\ifcase#1\or *\or \dagger\or \ddagger\or
%\mathsection\or \mathparagraph\or \|\or **\or \dagger\dagger
%\or \ddagger\ddagger\or \mathsection\mathsection
%\or \mathparagraph\mathparagraph \or \|\|\else\@ctrerr\fi}}
%\renewcommand{\thefootnote}{\fnsymbol{footnote}}
%\makeatother

\submitted{}
\title{The Surface Mass Density and Structure of the Outer Disk of NGC 628 \footnotemark[*]} 
\author{St\'ephane Herbert-Fort$^1$, Dennis Zaritsky$^1$, \\
Daniel Christlein$^2$, Sheila J.\ Kannappan$^{3}$}
\vspace{0.15cm}
\affil{
$^1$University of Arizona/Steward Observatory, 933 N Cherry Avenue, Tucson, AZ 85721\\
 (email: shf@as.arizona.edu) \\
$^2$Max-Planck-Institut f\"ur Astrophysik,
Karl-Schwarzschild-Str. 1, 85748 Garching, Germany\\
$^3$University of North Carolina/Department of Physics and Astronomy, 290 Phillips Hall CB 3255, \\Chapel Hill, NC 27599\\
}

\begin{abstract}

We study the kinematics of GALEX-selected \Ha\ knots in the outer disk (beyond 
$R_{25}$) of NGC 628 (M74), a galaxy representative of large, undisturbed, 
extended UV (Type 1 XUV) disks.  
Our spectroscopic target sample of 235 of the bluest UV knots surrounding NGC 628 yielded 15 \Ha\ 
detections ($6\%$), roughly the number expected given the different mean ages of the two populations.  
The measured vertical 
velocity dispersion of the \Ha\ knots between $1 - 1.8 R_{25}$ ($13.5 - 23.2$ kpc) is 
$< 11$ \kms.  We assume that the \Ha\ knots trace an 
`intermediate' vertical mass density distribution (between the isothermal sech$^2(z)$ and exponential 
distributions) with a constant scaleheight across the outer disk ($h_z$ = 700 pc) and 
estimate a total surface mass density of 7.5 $M_{\sun}$ pc$^{-2}$.  This surface mass 
density can be accounted for by the observed gas and stars in the outer disk (little or no dark matter 
in the disk is required).  
The vertical velocity dispersion of the outer disk \Ha\ knots nearly matches that measured 
from older planetary nebulae near the outskirts of the optical disk by Herrmann et al., 
suggesting a low level of scattering in the outer disk.  
A dynamically cold stellar component extending 
nearly twice as far as the traditional optical disk poses interesting constraints on the 
accretion history of the galaxy.
\end{abstract}
\footnotetext[*]{This paper includes data gathered with the 6.5 meter Magellan Telescopes located at Las Campanas Observatory, Chile.}

% for title footnote (defined above -- reset here)
\renewcommand{\thefootnote}{\arabic{footnote}}

\section{Introduction}

The outskirts of a galaxy are expected to host signatures of disk formation and 
hierarchical accretion because of the long dynamical times at these radii.  
\cite{TO92} argued that the 
thinness and coldness of inner galactic disks (inside the optical radius, $R_{25}$) 
posed significant problems for hierarchical models of galaxy formation.  
Qualitatively, these concerns become more pronounced if one can establish that cold 
galactic disks extend to even larger radii.

Various studies have revisited the cold disk problem, generally finding that accretion 
events are less destructive than originally envisioned.  Most recently, 
\cite{Kazan09} studied the dynamical response of thin 
galactic disks to bombardment by cold dark matter substructure out to large radii 
in fully self-consistent, dissipationless 
$N$-body simulations.  They found that disks survive these bombardments, 
but that they produce 
considerable thickening and heating at all radii, substantial flaring, and an increase 
in the stellar surface density in the disk outskirts (the latter due to 
outward radial migration of old stars during the growth and 
redistribution of disk angular momentum).  
Observations of the dynamical state of outer stellar disks \citep{Christlein08} demonstrate 
that outer disks generally continue to obey the flat rotation curves of inner disks, with no increase 
in the in-plane velocity dispersion.  
Here we present the face-on kinematics of one nearby galaxy, NGC 628 (M74), 
and compare to both neutral hydrogen and existing stellar measurements at smaller radii.  
NGC 628 is a prototypical 
Grand Design spiral galaxy \citep[type SA(s)c, dynamical mass within the studied region 
$\sim3.3\sci{11} M_{\sun}$, assuming $v_{rot} = 200$ \kms\ out to $2.3 R_{25}$;][]{Thilker07, KB92}, 
and is by far the largest and most massive member of its small group 
(the brightest member after NGC 628, UGC1176, is $\sim4.5$ mag fainter and over 125 kpc away).  
NGC 628 shows a standard exponential optical light profile to $R_{25}$, with only a very slight 
possible downbending in the profile to $\sim1.3 R_{25}$, the extent of the deep 
optical observations \citep{Natali92}.  Using ultraviolet (UV) imaging, 
\cite{Thilker07} classified NGC 628 as a Type 1 extended UV (XUV) disk, due to 
the structured, UV-bright emission complexes seen in the outer disk.  
Because the stellar disk appears largely undisturbed (both in optical and UV 
imaging, and, as we will show, from the kinematics) and because NGC 628 
dominates the dynamics of its local environment, we treat 
it as representative of large, isolated spiral galaxies in the nearby universe.

The outer disks of galaxies have 
received much recent attention, both observational 
\citep{Thilker07, GildePaz07, ZC07, Christlein08, HF09, Trujillo09, Herrmann3}
and theoretical \citep{Bush08, Roskar08a, Roskar08b, Kazan09}.  
The surge in interest in outer disks 
has been fueled by recent ultraviolet (UV) observations of nearby disks with the 
GALEX satellite \citep{Martin05, Thilker07}.  Previously, \cite{Ferg98} had used deep \Ha\ 
imaging to discover star formation in an 
extended component around three nearby galaxies, yet the 
ubiquity of this component in other disks remained largely unrecognized until 
the UV observations.  

This neglect stemmed in part from the fact that \Ha\ traces a limited subpopulation 
of the outer disk, namely those regions with O and B stars. Even relatively young 
regions may lack OB stars because they are older than 10 Myr or because they 
simply did not form such massive stars \citep{Ferg98, HF09, PA08}.
As such, the ubiquity of outer disk star formation is somewhat concealed. These 
barriers are removed with UV observations and we now know that many nearby 
galaxies host young outer disk stellar populations \citep[$>30\%$;][]{Thilker07, ZC07}.  
While galactic interactions can dramatically increase the level of outer disk star formation 
\citep[see the well-known cases of M83, NGC 4625 and M94;][]{Thilker05, GildePaz05, Trujillo09}, 
even isolated galaxies show low-levels of ongoing star formation in their outer parts 
\citep{Ferg98, Christlein08, HF09}.
However, the \Ha\ knots provide the bright emission lines that make it possible 
to measure the kinematics we present here.

While the gaseous outer disk of NGC 628 has been well-studied, for example by \cite{KB92}, the 
stellar component near the outskirts of the disk remains poorly characterized.  
Observers have acquired deep broadband imaging of the outer disk of NGC 628 
\citep[e.g.][]{Natali92}, however the diffuse light becomes difficult to reliably trace 
fainter than $\sim28$ mag arcsec$^{-2}$ in $V$, or beyond $\sim1.3 R_{25}$.  
To sidestep the difficulty of obtaining detailed kinematics from low surface brightness emission, 
\cite{Herrmann3} use planetary 
nebulae (PNe) across the face of NGC 628 to trace the underlying stellar distribution and to 
estimate the total mass density of the disk using the kinematic approach of \cite{vdKruit88} 
(specifically, by measuring the vertical velocity dispersion of a tracer of the disk mass 
distribution to estimate the underlying surface mass density).  
In principle, the PNe provide an excellent approach to the problem, with large numbers of PNe 
available for analysis.  Unfortunately for our purposes, the \cite{Herrmann3} study of PNe in 
NGC 628 is mostly constrained to the inner disk (only two of their PNe are beyond $R_{25}$; 
their study of M83 and especially M94 yield more PNe in those outer disks, however those 
are interacting systems for which the kinematics are complicated by recent events).  
The current work represents the first to analyze the disk kinematics out to $1.8 R_{25}$ 
in NGC 628.  Nevertheless, the results from 
\cite{Herrmann3} at the largest radii ($\sim R_{25}$) 
provide an interesting comparison to the results found here, 
especially because the PNe trace a population of objects that is 
on average at least $100\times$ older than the 
\Ha\ knots considered here.

We use multiobject spectroscopy to search for \Ha\ emission 
associated with blue GALEX sources in the outer disk of NGC 628.  
We measure the dispersion of the \Ha\ knot 
vertical velocity distribution, estimate the surface density of the outer disk, and determine 
if the baryonic material can account for the inferred mass.  
We then compare the results from the young \Ha\ knots to those from the PNe and 
consider the implications on disk evolution from the two measurements.
Section 2 presents our sample selection, observations and data reductions.  
Section 3 presents our analysis, 
results and discussion, including our measurement of the vertical velocity dispersion 
of the \Ha\ knots and the surface mass density of the outer disk.  
Section 4 presents a summary and our conclusions.

\section{Sample Selection, Observations, and Data Reduction}

We selected our target sample from sources identified in GALEX imaging of NGC 628 by 
\cite{ZC07}.  We target `blue' sources with FUV $-$ NUV $< 1$ and NUV $< 25$, corresponding 
to the expected colors of star clusters with ages $< 360$ Myr.  
We observed two fields that cover the outer disk of NGC 628 with the Inamori Magellan 
Areal Camera and Spectrograph \citep[IMACS;][]{Dressler06} using the `short' 
f/2 camera mode and the 600 line grating centered at 7695 \AA, providing a spectral resolution of 
$\sim 0.6$ \AA\ pixel$^{-1}$, or $\sim 27$ \kms\ pixel$^{-1}$ near H$\alpha$, 
on August 2, 3, and 4, 2005 (UTD).  The IMACS field of view in the 
short camera mode is $\sim 27^\prime$ on a side, providing coverage out to $3 R_{25}$ 
\citep[$R_{25} = 5.2$ arcmin;][]{KB92}.  
The IMACS plate scale in our setup is  0.2 arcsec pixel$^{-1}$.
We cut 162 and 151 slits in the masks of 
the northeastern and southwestern fields, respectively, giving priority to the brightest 
GALEX sources with the bluest FUV $-$ NUV colors 
(clusters with \Ha\ emission are expected to have ages $< 10$ Myr, powered 
by massive stars that will make the clusters blue in the GALEX bands).  
The seeing remained below $1^{{\prime}{\prime}}$ during 
the observations.  We observed the northern field for a total of 18000 seconds, 
or 5 hours, and the southern field for half that time.  To be certain that no significant mask/sky 
alignment drifts occurred during the overall integrations, 
we observed neither field for more than 2700 seconds in a single exposure.  
We observed HeNeAr comparison arc lamps 
through each mask for wavelength calibration.

We reduce the data with the Carnegie Observatories System for MultiObject 
Spectroscopy (COSMOS; A.\ Oemler et al.) 
package and extract individual 2D rectified, wavelength-calibrated 
and sky-subtracted spectra in the wavelength range 
5600 -- 9000 \AA.  
Using standard IRAF routines\footnotemark[1],
we median combine the individual sky-subtracted spectra corresponding to each knot, 
rejecting pixel values $>3\sigma$ from the median of the pixel stack to eliminate cosmic rays.

\footnotetext[1]{IRAF is distributed 
by the National Optical Astronomy Observatories, which are operated by 
the Association of Universities for Research in Astronomy, Inc., under 
cooperative agreement with the National Science Foundation.}

\section{Analysis, Results \& Discussion}

\subsection{\Ha\ detections}

We begin our spectral analysis by visually inspecting all of the sky-subtracted 
and combined spectra to search for 
detectable \Ha\ flux from the targeted UV-bright knots.
For any cases with visually-detected \Ha\ emission,
we average the rows showing signal and 
measure the radial velocity centroid from this final 1D spectrum.  
The procedure is designed such that 
we may exclude any rows that have minor overlaps with neighboring 
spectra or contain any remaining artifacts.

We estimate the line centroid measurement errors to be $\sim8$ \kms\ 
from fits to the bright OH 6498.72 \AA\ sky line from numerous 
(pre sky-subtracted) spectra.  
This measurement error is a lower limit to the 
\Ha\ line measurement error because the sky line is typically brighter and cleaner than 
the \Ha\ detections in our sample.
We apply a constant $-28.4$ \kms\ 
heliocentric velocity correction to the radial velocities of the \Ha\ knots; 
the difference in the heliocentric correction 
between consecutive nights is $\sim 0.1$ \kms, or $\sim1\%$ of the 
typical line centroid measurement error. 

\begin{figure}[h]
\centerline{
\includegraphics[angle=90,width=3.5in]{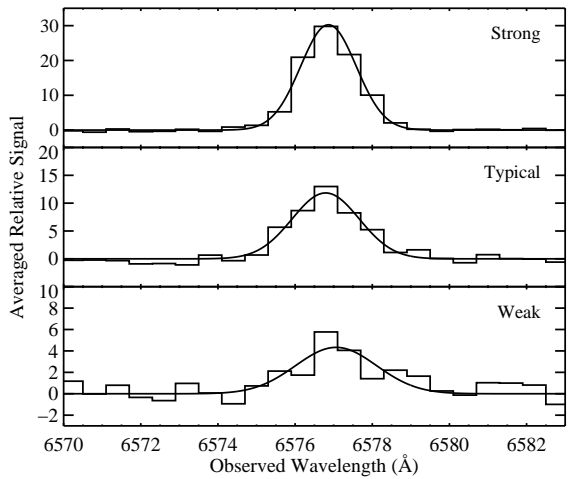}}
\caption{Three of our final 1D spectra, illustrating the range of strength of the \Ha\ detections.  Note the 
different $y$-axis ranges.  
Qualitatively, our final sample is composed of two `Strong', nine `Typical', and four `Weak' \Ha\ detections. 
\label{figspec}}
\end{figure}

\begin{figure}[h]
\centerline{
\includegraphics[angle=0,width=3.3in]{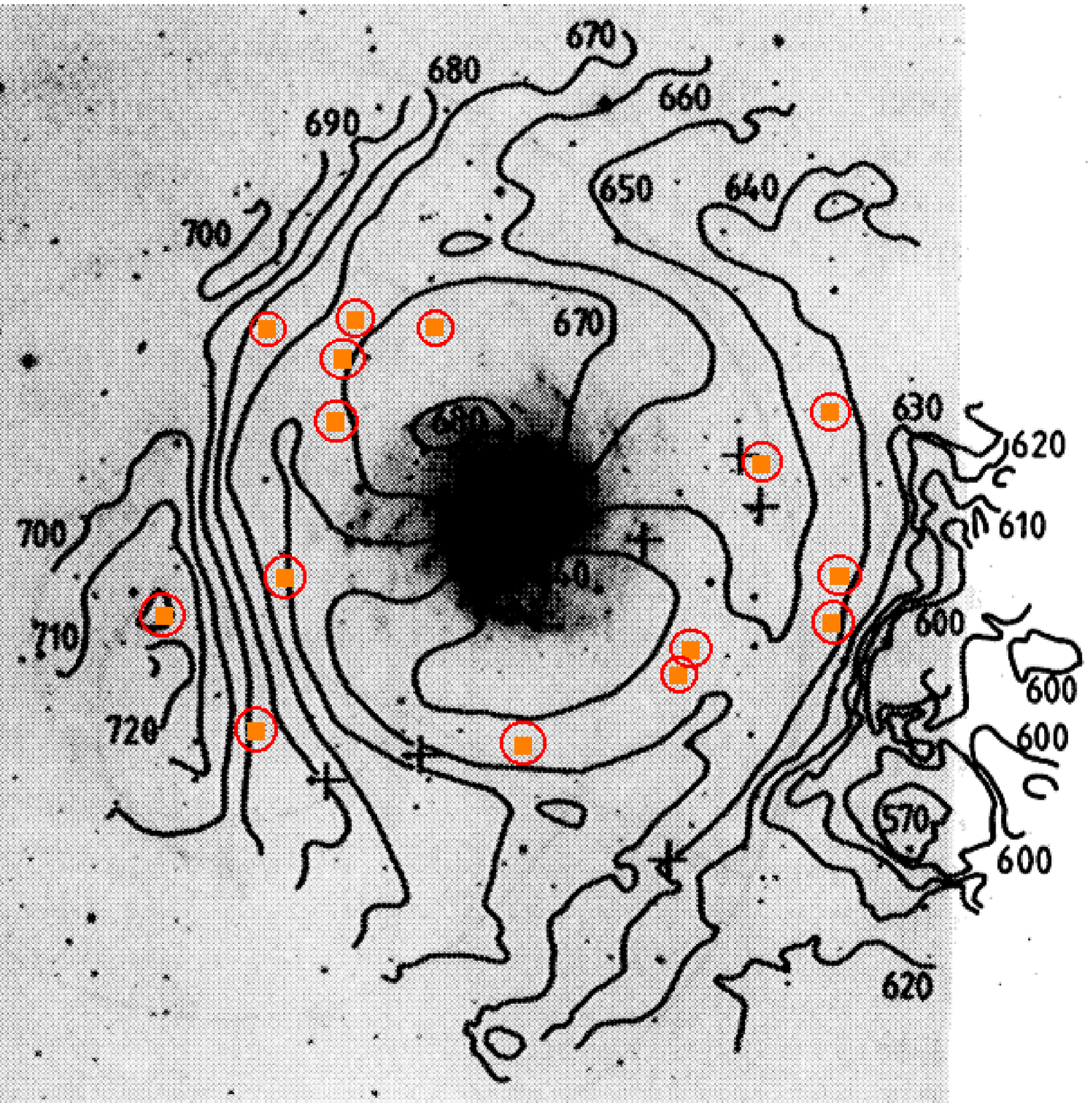}}
\caption{Figure 5 of \cite{KB92} showing the velocity field of the \ion{H}{1}-emitting gas disk surrounding NGC 628, with our 15 
\Ha\ detections overplotted as encircled filled squares.  The image is $\sim30^\prime$ on a side.  
The \Ha\ knots are found between $\sim 1$ and 1.8 $R_{25}$ (13.5 - 23.2 kpc) 
\label{N628_field}}
\end{figure}

Of the 313 slits on the two IMACS masks, $\sim25\%$ returned unusable 
spectra due to overlapping spectra or other artifacts in the data.  We therefore 
only search for \Ha\ emission in the remaining 235 spectra.  Fifteen of these 
spectra (or $\sim6\%$) yield detectable \Ha\ emission.  
Figure~\ref{figspec} presents a few of our final 1D spectra illustrating the range of 
strength of the \Ha\ detections.  
Our $6\%$ return is roughly what is 
expected given the difference in ages between the `blue' GALEX-detected knots 
targeted here ($<360$ Myr) and \Ha\ knots ($\sim10$ Myr) -- especially when 
considering that priority was given to the bluest and brightest GALEX knots 
(so most targeted GALEX knots are likely $<200$ Myr old).  
We plot the locations of these fifteen knots in Figure~\ref{N628_field} 
and list the relevant information in Table~\ref{tab:dets}.  
The physical scale of the individual \Ha\ knots (calculated from the extent of \Ha\ emission 
along the slit) range from 25 to 92 pc, with a 
median value of 67 pc \citep[adopting a distance of 8.6 Mpc to NGC 628, determined 
using the planetary nebula luminosity function;][]{Herrmann1}.

\begin{deluxetable}{cccccr}
\tablewidth{240pt}
\tablecaption{\Ha\ detections \label{tab:dets}}
\tabletypesize{\footnotesize}
\tablehead{\colhead{RA (J2000)} &\colhead{Dec (J2000)} &\colhead{$R/R_{25}$} & \colhead{$v_{helio}$} & \colhead{$v_{TH}$} & \colhead{$\Delta v$}}
\startdata
01:36:04.47  & +15:45:27.3  & 1.76 & 640 & 641 &   $-$1 \\
01:36:05.40  & +15:44:23.3  & 1.76 & 668 & 642 &  26 \\
01:36:06.53  & +15:49:37.0  & 1.72 & 652 & 647 &   5 \\
01:36:14.07  & +15:48:19.0  & 1.32 & 667 & 652 &  15 \\
01:36:20.94  & +15:43:33.5  & 1.17 & 644 & 643 &   1 \\
01:36:22.12  & +15:42:58.8  & 1.20 & 628 & 643 &  $-$15 \\
01:36:38.71  & +15:40:52.1  & 1.19 & 651 & 642 &   9 \\
01:36:51.02  & +15:52:03.6  & 1.06 & 680 & 684 &   $-$4 \\
01:36:59.23  & +15:52:13.6  & 1.29 & 657 & 670 & $-$13 \\
01:37:00.35  & +15:51:19.1  & 1.20 & 659 & 675 & $-$16 \\
01:37:00.94  & +15:49:49.7  & 1.04 & 658 & 674 & $-$16 \\
01:37:05.30  & +15:44:45.3  & 1.18 & 639 & 649 & $-$10 \\
01:37:09.28  & +15:41:14.0  & 1.70 & 704 & 679 &  25 \\
01:37:10.06  & +15:52:08.9  & 1.64 & 664 & 677 & $-$13 \\
01:37:17.86  & +15:43:51.0  & 1.79 & 656 & 711 & $-$55 \\
\enddata
\tablenotetext{*}{all velocities in \kms.  $v_{helio}$ refers to the heliocentric velocities of the \Ha\ knots, while $v_{TH}$ refers to THINGS heliocentric \ion{H}{1}-emitting gas velocities measured by \cite{Walter08}.  $\Delta v =  v_{helio} - v_{TH}$}
\end{deluxetable}

\subsection{\Ha\ knot velocities in the disk frame and $\sigma_z$}

Our present aim is to measure the vertical velocity dispersion  of 
the knots in the outer disk of NGC 628 ($\sigma_z$, with $z$ defined perpendicular to the 
disk plane).  Given the low inclination of NGC 628 
\citep[$i \sim 6^{\circ}$;][]{KB92}, the observed line-of-sight 
velocities are almost entirely due to the actual vertical velocities.  
However, as \cite{KB92} have shown, the outer gas 
disk of NGC 628 is chaotic and non-uniformly rotating.  
Whether this is also true for the outer stellar disk is unknown, and therefore 
whether we should use the previous orientation parameters obtained for the inner disk or
some other ones is unclear.  We therefore present 
results using various models for the disk orientation and kinematics, and accept the 
model with the lowest resulting velocity dispersion as the one most likely to be correct. 
Results of the following analysis will be discussed further in \S4.

We use a maximum likelihood approach to estimate the 
Gaussian parameters best describing the 
{\it{unbinned}} relative velocity ($\Delta v$) distributions.  
Specifically, we search a grid of mean and dispersion values 
(from $-50$ to $50$ \kms\ and $1$ to $100$ \kms, respectively) for the combination that yields the highest 
total probability (likelihood) of the Gaussian function, using the 
unbinned $\Delta v$ values as the independent variable.  
Our velocity dispersion estimates 
will provide only a single value for the radial range spanned by the knots ($\sim1 - 1.8 R_{25}$).  
We have insufficient detections to measure 
the velocity dispersion as a function of radius.  The following 
analysis will therefore consider the outer disk as a constant velocity dispersion, fixed-height component, 
and the values derived only represent the disk between $\sim1 - 1.8 R_{25}$.  
In support of this approach, note that 
the simulations of \cite{Kazan09} predict a roughly constant velocity dispersion in the outer disk.

We examine four models for the outer disk orientation and kinematics.  
First, we test a purely face-on disk by assuming a constant 656 \kms\ systemic 
velocity.  
This `model' yields a velocity dispersion of $18 \pm 5$ \kms.  
Second, we adopt the parameters of the inner neutral hydrogen gas disk of \cite{KB92} 
(inclination $i = 6.5^{\circ}$, major axis position angle $PA = 25^{\circ}$ and 
a mean velocity of 656 \kms) 
and a rotation velocity of 170 \kms\ \citep{Fathi07} and find a 
velocity dispersion of $19 \pm 5$ \kms.  Third, we adopt the two-component model 
of \cite{KB92} ($i = 13.5^{\circ}$ and $PA = 75^{\circ}$ 
beyond $1.2 R_{25}$, using the same inner disk values as the previous case 
for the knots found at $R < 1.2 R_{25}$) and 
find a velocity dispersion of $31 \pm 10$ \kms.  Finally, we adopt the neutral gas 
velocities from The \ion{H}{1} Nearby Galaxy Survey \citep[THINGS; ][]{Walter08} 
at the position of each knot and find a velocity dispersion of $19 \pm 6$ \kms.  
Figure~\ref{N628_field} shows the spatial distribution of the knots around the 
gas disk of NGC 628, and THINGS gas velocities at the positions of each 
knot are listed in Table~\ref{tab:dets}. 
This `model' tests for the possibility that the knots have low velocity dispersion 
around a complex velocity field defined by the \ion{H}{1}-emitting gas.  
Given the result that the velocity dispersion is similar to the simple disk models, 
we conclude that there is no evidence for highly complex kinematics.

The velocity dispersion estimates generally cluster around 20 \kms\ and 
agree to within 1 $\sigma$ (although the one estimate at $31 \pm 10$ \kms\ is 
just consistent with the others).  
The general agreement, independent of the model used, suggests that the dominant source 
of uncertainty is currently the statistical one associated with the limited sample size.  Because the 
fourth `model' is taken directly from the \ion{H}{1} data and produces a velocity dispersion as tight 
as do the other models, we adopt this model for the remainder of the paper.

\begin{figure}[h]
\centerline{
\includegraphics[angle=90,width=3.5in]{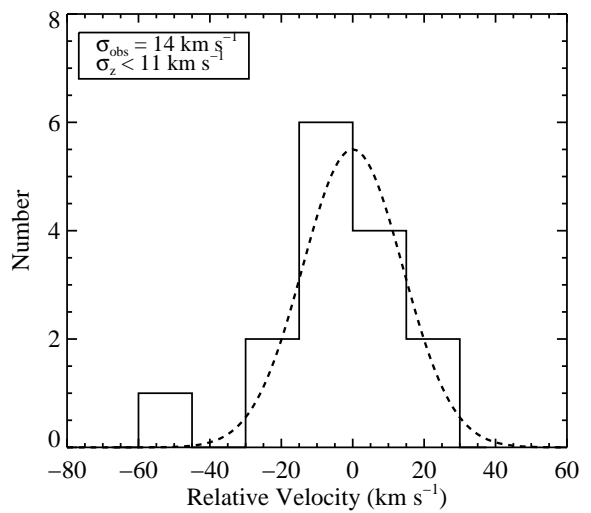}}
\caption{Distribution of \Ha\ knot vertical velocities, relative to THINGS \citep{Walter08} neutral 
gas measurements.  The outlier is the source detected on the high velocity cloud; 
we ignore this source in the analysis because its relation to the gas disk is uncertain.
\label{vsys}}
\end{figure}

Figure~\ref{vsys} shows the distribution of velocities relative to the \cite{Walter08} gas disk 
values (the fourth model), binned by 15 \kms.  
We removed a slight, but significant, mean velocity offset (17 \kms) when evaluating the 
velocity dispersion; Table~\ref{tab:dets} and Figure~\ref{vsys} include this correction.  
We attribute this shift to a velocity zero point difference in the 
various datasets although we were unable to directly identify the cause 
(Figure~\ref{N628_field} shows that the \Ha\ knots are well-distributed across the face of the disk; 
we therefore do not expect that the offset is due to an undersampling of the relative 
velocity distribution).  
When we exclude the knot at 01:37:17.86, +15:43:51.0 with 
$\Delta v =  v_{helio} - v_{TH} = -55$ \kms, which lies on the high velocity cloud 
(HVC) described in \cite{KB92}, we arrive at our best (lowest and most likely) 
estimate of the observed \Ha\ knot velocity dispersion, $14 \pm 4$ \kms.  Excluding 
the same knot from the other models does not significantly change their values.  
The Gaussian representing this final {\it{unbinned}} 
$\Delta v$ distribution is overplotted in Figure~\ref{vsys} 
(for presentation, the height of the Gaussian is determined by a fit to 
the binned $\Delta v$ distribution; the dispersion is the only value used in our 
analysis, however).

We now consider the degree to which our observational errors inflate the velocity 
dispersion measurement ($\sigma_{obs} = 14$ \kms).  
Subtracting our previous velocity error estimate (8 \kms) in quadrature from 
the measured velocity dispersion, we estimate that $\sigma_{z} < 11$ \kms.  
We emphasize that $\sigma_{z}$ is likely to be an upper limit to the vertical velocity dispersion 
because 1) the measurement uncertainty is larger than 8 \kms, 2)  the 
uncertainty in the reference \ion{H}{1}-emitting gas velocities was ignored, and 
3) we may have missed the correct model for the underlying disk kinematics.

Our low velocity dispersion measurement suggests a relatively undisturbed stellar disk, which at 
first seems to contrast with the \cite{KB92} observation of a chaotic outer gas disk.  The \cite{KB92} 
observation was strongly influenced by the HVCs and kinematic features at the largest radii, 
however--the bulk of the gas disk is in fact rather undisturbed (see Figure~\ref{N628_field}).  
Nevertheless, a difference between a relatively undisturbed stellar disk and a disturbed 
gas disk might be expected, considering the 
results of \cite{Thilker07} and \cite{Moffett09}, who find that Type 1 XUV disks 
(of which NGC 628 is a member) cover a wide range 
of galaxy masses, colors and morphologies, and that these disks may in fact be associated 
with minor interactions or accretion.  
The disturbed structure observed in the gas may 
result from an interaction event \citep{KB92} 
to which the stellar disk was less sensitive.  Such an event 
may have then triggered the Type 1 XUV disk that is observed today.

\subsection{Outer disk scale height $h_z$}

Assuming that the \Ha\ knot density has the same vertical distribution as 
the underlying mass density distribution of the disk, 
for an adopted scaleheight $h_z$ we constrain the surface mass density of the outer disk 
$\Sigma$ using our measurement of $\sigma_z$.  
We will return to discuss the validity of this assumption, 
although we do not have any alternative given our current data. 
The relevant equation \citep{vdKruit88} is 

\begin{equation}
\sigma_z^2 = K G \Sigma h_z,
\end{equation}

\noindent where $K = 1.7051\pi$ for an `intermediate' vertical mass density distribution  
(between the isothermal sech$^2(z)$ and exponential 
distributions) and $G$ is the gravitational constant.  
The relation was constructed by solving the Boltzmann and Poisson 
equations for plane parallel layers.  For a galaxy 
with a flat rotation curve, the plane parallel case provides a realistic model from 
which to estimate the surface mass density \citep{vdKruit88}.  
Because \cite{Christlein08} found mostly flat rotation curves in a sample 
of 17 nearby outer disks, we expect the plane parallel framework to remain valid in the outer regions 
(we further assume a constant scale height over the radial range of the knots).  
Corrections to the relation from a dark halo term are negligibly small 
at low $z$ \citep{vdKruit88}; however, in the outer disk the 
baryonic disk contribution to the mass budget presumably drops more rapidly with 
radius than does the dark matter contribution, and so the dark matter 
may in fact not be negligible in the outer disk.  In the following section we 
show that the \cite{vdKruit88} relation is indeed appropriate at these radii 
(we find that no dark matter in the disk is required to explain existing observations, 
even when assuming a rather large value for $h_z$).

Because NGC 628 is nearly face-on, we must estimate the scale height $h_z$ 
of the tracer particles (the knots) indirectly.
\cite{Herrmann3} argue, based on stability arguments and the observed 
velocity dispersion of PNe, that the scaleheight near the edge of the region 
they probe ($\sim11$ kpc) must be $>400$ pc.  
The vertical disk flaring suggested by the PNe 
analysis resembles the known flaring of gas disks in their outer regions, 
with neutral hydrogen gas scale heights near $\sim700$ pc at the $R/R_{25}$ distances considered here 
\citep{Merrifield92, Corbelli93, Olling96}.
Because the \Ha\ knots are young and because they lie at radii larger than those probed by the PNe, 
we adopt a scale height for the knots in line with that of the 
flaring gas, $h_z = 700$ pc at 1.3$R_{25}$, the median radius of our 
\Ha\ detections.  Recall however that we treat the outer disk as a 
fixed-height component.  In reality, $h_z$ may range between 
$\sim400$ pc and $\sim1000$ pc over the radii covered by our detections.  
Assuming a constant 700 pc over the entire outer disk therefore 
implies a large uncertainty in the final outer disk mass, which we assume to 
be $\sim20\%$.  This level of uncertainty will not significantly affect our conclusions, however.  
Finally, we note that our assumption that the \Ha\ knots trace the same mass distribution as the outer disk 
gas is different from what one would expect in the inner disk.  
In the inner disk, molecular clouds (and their \Ha\ knots) have a smaller vertical scaleheight than 
the mass, which is dominated by the stellar component.  In the outer disk, however, 
where the dominant baryonic 
component is the gas (as we show below), our assumption is much more appropriate.  
Whether the \Ha\ knots trace the neutral gas is more uncertain, although where examined
in detail the correspondence appears good \citep{Thilker05,HF09}.
In view of this, a large scaleheight for the outer stellar disk should be expected.  
We caution however that if a significant amount of mass extends to heights larger than 
the 700 pc scaleheight assumed for the knots, we are not sensitive to it with the \Ha\ knot measurement.

%\newpage
\subsection{Outer disk mass density}

We now estimate the mass density of the outer disk in order to determine 
the relative contributions of stars and gas to the total outer disk mass. 
Using $\sigma_z = 11$ \kms\ and 
our adopted $h_z = 700$ pc, we calculate that the outer disk surface 
mass density  $\Sigma$ is 7.5 M$_{\sun}$ pc$^{-2}$.  
The total mass between $1 - 1.8 R_{25}$ is then $\sim8.4 \sci 9 M_{\sun}$.

We now use results from deep optical imaging of the outer regions of 
NGC 628 \cite[$\mu_V \sim 27$ mag arcsec$^{-2}$ at 1.3 $R_{25}$;][]{Natali92} 
to estimate that the mass-to-light ratio $\Upsilon$ of the outer disk is $\sim18$.   
The outer disk mass is therefore not primarily in the form of normal stars.  
For comparison, \cite{Herrmann3} 
found that the PNe velocity dispersions imply $\Upsilon = 1.4$ for the inner 
disk.  
Reconciling the outer and inner disk $\Upsilon$ values 
would require a factor of 10 changes in $\sigma_z$ or $\Sigma$, 
neither of which is that poorly constrained.  Therefore, as one might expect, 
we detect a significantly higher $\Upsilon$ in the outer regions.  
If we adopt $\Upsilon = 3$ as an upper limit for an old outer disk stellar 
population \citep{Bell01}, 
we find that the dark material in the outer disk must be at least 
five times more massive than the stellar component.
In order to match the $\mu_V \sim 27$ mag arcsec$^{-2}$ observation, an outer 
disk stellar population with $\Upsilon < 3$ would require $\Sigma_\star < 1.3 M_{\sun}$ pc$^{-2}$ 
and $M_\star < 1.4 \sci 9 M_{\sun}$.  This leaves $> 7 \sci 9 M_{\sun}$ of dark material 
to be accounted for. 

We now consider the neutral gas component of the outer disk and whether it 
can account for the optically non-luminous mass.  
Using the outer disk \ion{H}{1}-emitting gas mass from \cite{KB92} (over the relevant radial range), 
$\sim6 \sci 9 M_{\sun}$, 
we estimate the total outer disk baryonic mass (atomic gas $+$ stars) 
to be between $\sim6 \sci 9 - 7.4 \sci 9 M_{\sun}$.  
Comparing this total baryonic mass to the total mass required 
by the \Ha\ knot velocity dispersion ($\sim8.4 \sci 9 M_{\sun}$), we find that the 
baryonic matter (mainly the \ion{H}{1}-emitting gas) 
can account for nearly all of the measured outer disk mass 
(note that we have ignored any molecular gas contribution).  
Our assumption that the dark matter contribution to the disk 
is negligible at large radii appears justified, as does our claim that the gaseous 
component dominates the baryon budget.

\subsection{Star formation history constraints}

A separate intriguing question is whether star formation has occurred at 
the current rate for the entire lifetime of this galaxy. The result of such a 
stellar component could, in principle, violate the surface brightness or surface 
mass density measurements. Independent of the surface brightness measurement 
we can simply calculate whether the integrated star formation results in a mass 
density in conflict with the measurement.  
Using deep \Ha\ imaging of three nearby late-type spiral galaxies (one of which is the current 
focus; NGC 628, NGC 1058 and NGC 6946), 
\cite{Ferg98} found their outer disk star formation rate densities to be between 
$\sim0.01 - 0.05 M_{\sun}$ pc$^{-2}$ Gyr$^{-1}$.  
For the following exercise, we adopt a SFR density of 
$0.03 M_{\sun}$ pc$^{-2}$ Gyr$^{-1}$ for the outer disk of NGC 628.
For comparison to the \cite{Ferg98} SFR densities, \cite{Trujillo09} used GALEX 
FUV imaging to measure the SFR density 
in the outer disk of M94 and found $\sim 0.4 M_{\sun}$ pc$^{-2}$ Gyr$^{-1}$.  
The star formation in the outer regions of M94 is unusually 
large however, producing an anti-truncated optical disk light profile \citep[NGC 628 
shows a standard exponential optical light profile to $R_{25}$, with only a very slight 
possible downbending in the profile to $\sim1.3 R_{25}$, the extent of the deep 
optical observations;][]{Natali92}.  
For NGC 628, assuming uniformly-distributed constant star formation over a Hubble time 
leads to an expected stellar outer disk mass of $5 \sci 8 M_{\sun}$ ($0.45 M_{\sun}$ pc$^{-2}$),  
well within our measured mass of $8.4 \sci 9 M_{\sun}$ 
and within the stellar mass inferred from the measured surface brightness 
($M_\star < 1.4 \sci 9 M_{\sun}$ for a population with $\Upsilon < 3$).  
We conclude that the outer disk may have been forming stars for the entire lifetime 
of the galaxy at approximately the current star formation rate.

\subsection{Similarities between the young and old components}

We now compare the outer disk velocity dispersions of the young ($<10$ Myr) and old 
(100 Myr $-$ 10 Gyr) populations, 
traced by the \Ha\ knots and PNe \citep{Herrmann3}, respectively, and find little 
difference ($\sim11$ \kms\ vs.\ $\sim12$ \kms; a small difference may be allowed 
if we treat the \Ha\ velocity dispersion as a true upper limit, however).  Because the difference is 
unlikely to be much larger than the $1\sigma$ error of the measurement, 
we cannot say that there is a 
significant difference in the kinematic or structural properties between the 
young and old outer disk populations.

A powerful tracer of the evolution of stellar populations
of different ages is the increase of velocity dispersion with
age, or, alternatively, the varying asymmetric drift of various
populations. The observed low $\sigma_z$ of the \Ha\ knots
immediately suggests that the degree of asymmetric drift will
be small.  If one considers that some fraction of the observed
value must reflect the intrinsic dispersion of the gas, then
the asymmetric drift will be even smaller. The other population
that has been measured, PNe, also has a low velocity dispersion
and the inferred asymmetric drift is also relatively small \citep[$<15$
\kms;][]{Herrmann3}. With the current measurements we
cannot conclude that there has been any increase in the
velocity dispersions between the very young \Ha\ knots
and the older PNe (we comment further on this issue in the following section).  
It is evident that such an investigation will
require uncertainties in the velocity dispersions better than $\sim1$ \kms.

\section{Summary and Conclusions}

We have measured the kinematics of \Ha\ knots in the outer disk of NGC 628.  
We find the stellar disk (traced by the \Ha\ knots) 
to have a low velocity dispersion, suggesting an undisturbed extended stellar disk.  
NGC 628 
differs from the better-known outer disks of M83 and M94, which were likely 
accentuated by recent disturbances from neighbors, but 
resembles (both in optical and kinematic profiles) the outer disks detected in other relatively 
undisturbed nearby edge-on disks by \cite{Christlein08}, supporting the idea 
that outer disk star formation can be a low-level and ongoing 
phenomenon in isolated galaxies.

We find $\sigma_z$ of the \Ha\ knots to be $< 11$ \kms\ between $1 - 1.8 R_{25}$ ($13.5 - 23.2$ kpc).  
We adopt a scaleheight similar to the known flaring gas profiles of outer disks 
($h_z = 700$ pc) and estimate a mass density $\Sigma = 7.5$ M$_{\sun}$ pc$^{-2}$ that 
can be entirely explained by the observed gas and stars in the outer disk.  
If the \Ha\--hosting disk is actually much thinner than the flaring gas disk, 
more dark matter in the outer regions would be allowed.  
Assuming that outer disk star formation has been going for a Hubble time 
does not violate either the surface brightness nor surface mass constraints.  
The high incidence of outer disks \citep[c.f.][]{Christlein08} suggests that the star formation is 
not a rare phenomenon -- here we show that current limits cannot exclude long-lived 
outer disk star formation.

Finally, the velocity dispersion of 
PNe towards the outer disk of NGC 628 \citep{Herrmann3} 
is nearly the same as that of the \Ha\ knots (the discrepancy grows slightly when 
considering the dispersion of the \Ha\ knots as a strict upper limit, 
although any plausible difference remains small). 
This can result either if the 
PN population is rather young (so that scattering has not had a chance to enlarge the 
dispersion significantly), if there is very little scattering, or if scattering occurs 
primarily in a single (or few) discrete events that occurred prior to the creation of 
the bulk of the PNe.  
We do not expect the outer disk to be exclusive young (see above) 
and we do expect some level of scattering in outer disks 
(if not from the classical spiral arms and molecular 
clouds of inner disks, then perhaps from satellites and dark halo substructure).  
The solution may be found in the \cite{Kazan09} simulations, 
which show that outer disk heating is dominated by
the most massive infall event of halo substructure onto the disk, so 
that stellar populations of different ages do not necessarily have different velocity
dispersions.  A larger sample of outer disk kinematic measurements could be used to 
constrain the rate and impact of such infall events.

\section{Acknowledgments}

This work made use of THINGS, `The \ion{H}{1} Nearby Galaxy Survey' \citep{Walter08}.  
DZ and SHF were supported by 
NSF AST-0907771 and NASA LTSA NNG05GE82G.  
SJK was supported by NSF AST-0401547.


\newpage
\begin{thebibliography}{}

\bibitem[Bell \& de Jong (2001)]{Bell01} % old pop M/L
Bell, E.F. \& de Jong, R.S.  2001, ApJ, 550, 212

\bibitem[Bessell et al.\ (1986)]{Bessell86} % old stars show same structure and systemic velocity as HI and young stars in LMC
Bessell, M.S., Freeman, K.C., Wood, P.R. 1986, ApJ, 310, 710

\bibitem[Bush et al.\ (2008)]{Bush08}	% sims of XUV disk formation with density threshold 
Bush, S.J., Cox, T.J., Hernquist, L., Thilker, D., Younger, J.D.  2008, ApJL, 683, 13

\bibitem[Christlein \& Zaritsky (2008)]{Christlein08}  % flat rotation curves in outer disks, in general
Christlein, D., Zaritsky, D. 2008, ApJ, 680, 1053

\bibitem[Corbelli \& Salpeter (1993)]{Corbelli93}  % outer HI h_z likely > 500 pc M33, N3198
Corbelli, E., Salpeter, E.E.  1993, ApJ, 419, 104

\bibitem[Dressler et al.\ (2006)]{Dressler06}  % IMACS (once in use) ref
Dressler, A., Hare, T., Bigelow, B.C., Osip, D.J.  2006, SPIE, 6269, 13

%original IMACS ref?:  Bigelow et al.\ (1998)

\bibitem[de Vaucouleurs et al.\ (1976)]{dV76}  % R_25 of 5.2 arcmin (N628)
de Vaucouleurs, G., de Vaucouleurs, A., \& Corwin, H.G. 1976, 
Second Reference Catalogue of Bright Galaxies (Austin : Univ.\ of Texas Press)


\bibitem[Fathi et al.\ (2007)]{Fathi07} % Vrot = 170 from SAURON Ha obs inner disk
Fathi, K., Beckman, J.E., Zurita, A., Rela\~no, M., Knapen, J. H., Daigle, O., Hernandez, O., Carignan, C.  2007, A\&A, 466, 905

\bibitem[Ferguson et al.\ (1998)]{Ferg98} % outer disk Ha in N628, outer disk SFR ~0.04 Msun/pc2/Gyr
Ferguson, A.M.N., Wyse, R.F.G., Gallagher, J.S., Hunter, D.A.  1998, ApJ, 506, 19

\bibitem[Gil de Paz et al.\ (2005)]{GildePaz05}  % XUV disk of NGC 4625
Gil de Paz, A. et al. 2005, ApJL, 627, 29

\bibitem[Gil de Paz et al.\ (2007)]{GildePaz07}  % GALEX atlas of XUV disks (APJS)
Gil de Paz, A. et al. 2007, ApJS, 173, 185

\bibitem[Herbert-Fort et al.\ (2009)]{HF09}  % LBT N3184 outer disk 
Herbert-Fort, S.\ et al.\ 2009, ApJ, 700, 1977

\bibitem[Herrmann et al.\ (2008)]{Herrmann1}  % PNLF distance to N628 = 8.6 Mpc
Herrmann, K.A., Ciardullo, R., Feldmeier, J.J., Vinciguerra, M.  2008, ApJ, 683, 630

\bibitem[Herrmann et al.\ (2009)]{Herrmann3}  % PNe kinematics and disk masses
Herrmann, K.A., Ciardullo, R.  2009, ApJ, in press

\bibitem[Kazantzidis et al.\ (2009)]{Kazan09} % halo subtructure heats outer disks
Kazantzidis, S., Zentner, A.R., Kravtsov, A.V., Bullock, J.S., Debattista, V.P.  
2009, ApJ, 700, 1896

\bibitem[Kamphuis \& Briggs (1992)]{KB92}  %  wide field HI map of N628 -- chaotic outer disk
Kamphuis, J. \& Briggs, F.  1992, A\&A, 253, 335

\bibitem[Larson \& Tinsley (1978)]{LT78} % used by Natali92 -- two-color plot vs. 'SFR' 
Larson, R.B., Tinsley, B.M.  1978, ApJ, 219, 46

\bibitem[Martin et al.\ (2005)]{Martin05}	% GALEX ref
Martin, D.C.\ et al.\ 2005, ApJ, 619, 1L

\bibitem[Merrifield (1992)]{Merrifield92} % MW outer HI h_z > 500pc
Merrifield, M.R. 1992, AJ, 103, 1552

\bibitem[Moffett et al.\ (2009)]{Moffett09}  % Type 1 XUV disks may be associated with previous interactions
Moffett, A.J., Kannappan, S.J., Laine, S., Wei, L.H., Baker, A.J., Impey, C.D.  2009, arXiv0908.4232

\bibitem[Natali et al.\ (1992)]{Natali92}	% deep UBVRI imaging of N628 -- outer disk SB in V
Natali, G., Pedichini, F., Righini, M.  1992, A\&A, 256, 79

\bibitem[Olling (1996)]{Olling96}  % outer HI h_z  ~800 pc from N4244
Olling, R.P. 1996, AJ, 112, 457

\bibitem[Pflamm-Altenburg \& Kroupa (2008)]{PA08}  % clustered SF, local IGIMF theory
Pflamm-Altenburg, J., Kroupa, P.  2008, Nature, 455, 641

\bibitem[Ro$\check{\rm{s}}$kar et al.\ (2008a)]{Roskar08a}  % Radial migration outer disks
Ro$\check{\rm{s}}$kar, R., Debattista, V.P., Stinson, G.S., Quinn, T.R., 
Kaufmann, T., Wadsley, J.  2008, ApJL, 675, 65

\bibitem[Ro$\check{\rm{s}}$kar et al.\ (2008b)]{Roskar08b}
Ro$\check{\rm{s}}$kar, R., Debattista, V.P., Quinn, T.R., Stinson, G.S., Wadsley, J.  
2008, ApJL, 684, 79

\bibitem[Thilker et al.\ (2005)]{Thilker05}  % M83 GALEX letter
Thilker, D.A.\ et al.  2005, ApJL, 619, 79

\bibitem[Thilker et al.\ (2007)]{Thilker07}  % GALEX XUV disks
Thilker, D.A. et al.\ 2007, ApJS, 173, 538

\bibitem[Toomre (1964)]{Toomre64}  %  disk stability against 
Toomre, A. 1964, ApJ, 139, 1217

\bibitem[Toth \& Ostriker (1992)]{TO92} % disk heating by infall, stability arguments
Toth, G., Ostriker, J.P.  1992, ApJ, 389, 5

\bibitem[Trujillo et al.\ (2009)]{Trujillo09}  % M94 outer disk SFR (0.15 Msun/yr)
Trujillo, I., Martinez-Valpuesta, I., Mart\'inez-Delgado, D., Pe\~narrubia, J., Gabany, R.J., Pohlen, M.  2009, ApJ, 704, 618

\bibitem[van der Kruit (1988)]{vdKruit88}  % equation for dispersion --> surf mass density
van der Kruit, P.C.  1988, A\&A, 192, 117

\bibitem[Walter et al.\ (2008)]{Walter08}  % THINGS paper
Walter, F., Brinks, E., de Blok, W.J.G., Bigiel, F., Kennicutt, R.C., Thornley, M.D., Leroy, A.  2008, AJ, 136, 2563

\bibitem[Zaritsky \& Christlein (2007)]{ZC07}  % Dennis' GALEX paper -- sample selection
Zaritsky, D., Christlein, D.  2007, AJ, 134, 135

\end{thebibliography}
\end{document}